\documentstyle[aps,eqsecnum,preprint,floats,epsf,epsfig]{revtex}
\begin{document}
\def\be{\begin{eqnarray}}
\def\en{\end{eqnarray}}
\def\non{\nonumber}
\def\la{\langle}
\def\ra{\rangle}
\def\nc{N_c^{\rm eff}}
\def\vp{\varepsilon}
\def\a{{\cal A}}
\def\B{{\cal B}}
\def\c{{\cal C}}
\def\d{{\cal D}}
\def\e{{\cal E}}
\def\p{{\cal P}}
\def\t{{\cal T}}
\def\up{\uparrow}
\def\dw{\downarrow}
\def\vma{{_{V-A}}}
\def\vpa{{_{V+A}}}
\def\smp{{_{S-P}}}
\def\spp{{_{S+P}}}
\def\J{{J/\psi}}
\def\ov{\overline}
\def\Lqcd{{\Lambda_{\rm QCD}}}
\def\pr{{\sl Phys. Rev.}~}
\def\prl{{\sl Phys. Rev. Lett.}~}
\def\pl{{\sl Phys. Lett.}~}
\def\np{{\sl Nucl. Phys.}~}
\def\zp{{\sl Z. Phys.}~}
\def\lsim{ {\ \lower-1.2pt\vbox{\hbox{\rlap{$<$}\lower5pt\vbox{\hbox{$\sim$}
}}}\ } }
\def\gsim{ {\ \lower-1.2pt\vbox{\hbox{\rlap{$>$}\lower5pt\vbox{\hbox{$\sim$}
}}}\ } }

\font\el=cmbx10 scaled \magstep2{\obeylines\hfill December, 2002}

\vskip 1.5 cm

\centerline{\large\bf Hadronic $D$ Decays Involving Scalar Mesons}
\bigskip
\centerline{\bf Hai-Yang Cheng}
\medskip
\centerline{Institute of Physics, Academia Sinica}
\centerline{Taipei, Taiwan 115, Republic of China}
\medskip

\bigskip
\bigskip
\centerline{\bf Abstract}
\bigskip
{\small The nonleptonic weak decays of charmed mesons into a
scalar meson and a pseudoscalar meson are studied. The scalar
mesons under consideration are $\sigma$ [or $f_0(600)$], $\kappa$,
$f_0(980)$, $a_0(980)$ and $K^*_0(1430)$. A consistent picture
provided by the data suggests that the light scalars below or near
1 GeV form an SU(3) flavor nonet and are predominately the
$q^2\bar q^2$ states, while the scalar mesons above 1 GeV can be
described as a $q\bar q$ nonet. Hence, we designate $q^2\bar q^2$
to $\sigma,~\kappa,~a_0(980),~f_0(980)$ and $q\bar q$ to $K^*_0$.
Sizable weak annihilation contributions induced from final-state
interactions are essential for understanding the data. Except for
the Cabibbo doubly suppressed channel $D^+\to f_0K^+$, the data of
$D\to\sigma\pi,~ f_0\pi,~f_0K,~K^*_0\pi$ can be accommodated in
the generalized factorization approach. However, the predicted
rates for $D\to a_0\pi,~a_0K$ are too small by one to two orders
of magnitude when compared with the preliminary measurements.
Whether or not one can differentiate between the two-quark and
four-quark pictures for the $f_0(980)$ produced in the hadronic
charm decays depends on the isoscalar $f_0\!-\!\sigma$ mixing
angle in the $q\bar q$ model.

}

\pagebreak

\section{Introduction}
There are two essential ingredients for understanding the hadronic
decays of charmed mesons. First, the nonfactorizable contributions
to the internal $W$-emission amplitude, which is naively expected
to be color suppressed, is very sizable. Second, final-state
interactions (FSIs) play an essential role. The nonfactorizable
corrections to nonleptonic charm decays will compensate the Fierz
terms appearing in the factorization approach to render the
naively color-suppressed modes no longer color suppressed. The
weak annihilation ($W$-exchange or $W$-annihilation) amplitude
receives long-distance contributions  via inelastic final-state
interactions from the leading tree or color-suppressed amplitude.
As a consequence, weak annihilation has a sizable magnitude
comparable to the color-suppressed internal $W$-emission with a
large phase relative to the tree amplitude. A well known example
is the decay $D^0\to\ov K^0\phi$ which proceeds only through the
$W$-exchange process. Even in the absence of the short-distance
$W$-exchange contribution, rescattering effects required by
unitarity can produce this reaction \cite{Donoghue}. Then it was
shown in \cite{CC87} that this rescattering diagram belongs to the
generic $W$-exchange topology.

There exist several different forms of FSIs: elastic scattering
and inelastic scattering such as quark exchange, resonance
formation,$\cdots$, etc. The resonance formation of FSIs via
$q\bar q$ resonances is probably the most important one to
hadronic charm decays owing to the existence of an abundant
spectrum of resonances known to exist at energies close to the
mass of the charmed meson. Since FSIs are nonperturbative in
nature, in principle it is notoriously difficult to calculate
their effects. Nevertheless, most of the properties of resonances
follow from unitarity alone, without regard to the dynamical
mechanism that produces the resonance \cite{Zen,Weinberg}.
Consequently, the effect of resonance-induced FSIs can be
described in a model-independent manner in terms of the masses and
decay widths of the nearby resonances (for details, see e.g.
\cite{a1a2charm}).

It has been established sometime ago that a least
model-independent analysis of heavy meson decays can be carried
out in the so-called quark-diagram approach \cite{Chau,CC86,CC87}.
Based on SU(3) flavor symmetry, this model-independent analysis
enables us to extract the topological quark-graph amplitudes for
$D\to PP, VP$ decays and see the relative importance of different
underlying decay mechanisms. From this analysis one can learn the
importance of the weak annihilation amplitude and a nontrivial
phase between tree and color-suppressed amplitudes \cite{Rosner}.

In the present paper we shall study the nonleptonic decays of
charmed mesons into a pseudoscalar meson and a scalar meson. Light
scalar mesons are traditionally studied in low energy $S$-wave
$\pi\pi$, $K\pi$ and $K\ov K$ scattering experiments and in $p\bar
p$ and $Z\bar N$ annihilations. Thanks to the powerful Dalitz plot
analysis technique, many scalar meson production measurements in
charm decays are now available from the dedicated experiments
conducted at CLEO, E791, FOCUS, and BaBar. Hence the study of
three-body decays of charmed mesons opens a new avenue to the
understanding of the light scalar meson spectroscopy.
Specifically, the decays $D\to f_0\pi(K)$, $D\to a_0\pi(K)$, $D\to
\ov K^*_0\pi$ and $D^+\to \sigma\pi^+$ have been observed.
Moreover, in some three-body decays of charmed mesons, the
intermediate scalar meson accounts for the main contribution to
the total decay rate. For example, $D_s^+\to f_0(980)\pi^+$ and
$D_s^+\to f_0(1370)\pi^+$ account for almost 90\% of the
$D^+_s\to\pi^+\pi^+\pi^-$ rate \cite{E791}, while about half of
the total decay rate of $D^+\to\pi^+\pi^+\pi^-$ comes from
$D^+\to\sigma\pi^+$ \cite{E791}.

The study of $D\to SP$ is very similar to $D\to PP$ except for the
fact that the quark structure of the scalar mesons, especially
$f_0(980)$ and $a_0(980)$, is still not clear. A consistent
picture provided by the data implies that light scalar mesons
below or near 1 GeV can be described by the $q^2\bar q^2$ states,
while scalars above 1 GeV will form a conventional $q\bar q$
nonet. Another salient feature is that the decay constant of the
scalar meson is either zero or very small. We shall see later that
final-state interactions are essential for understanding the $D\to
SP$ data. It is hoped that through the study of $D\to SP$, old
puzzles related to the internal structure and related parameters,
e.g. the masses and widths, of light scalar mesons can receive new
understanding.

This work is organized as follows. In Sec. II we summarize the
experimental measurements of $D\to SP$ decays and emphasize that
many results are still preliminary. We then discuss the various
properties of the scalar mesons in Sec. III, for example, the
quark structure, the decay constants and the form factors. Sec. IV
is devoted to the quark-diagram scheme and its implication for
final-state interactions. We analyze the $D\to SP$ data in Sec. V
based on the generalized factorization approach in conjunction
with FSIs. Conclusion is made in Sec. VI.

\section{Experimental status}
It is known that three-body decays of heavy mesons provide a rich
laboratory for studying the intermediate state resonances. The
Dalitz plot analysis is a very useful technique for this purpose.
We are interested in $D\to SP$ ($S$: scalar meson, $P$:
pseudoscalar meson) decays extracted from the three-body decays of
charmed mesons. Some recent results (many being preliminary) are
available from E791 \cite{E791}, CLEO \cite{CLEO}, FOCUS
\cite{FOCUS} and BaBar \cite{BaBar}. The $0^{+}$ scalar mesons
that have been studied in charm decays are $\sigma(500)$ [or
$f_0(600)$], $f_0(980)$, $f_0(1370)$, $a_0(980)$, $a_0(1450)$,
$\kappa$ and $K^*_0(1430)$. The results of various experiments are
summarized in Table I where the product of $\B(D\to SP)$ and
$\B(S\to P_1P_2)$ is listed. In order to extract the branching
ratios for $D\to f_0 P$, we use the value of $\Gamma(f_0\to
\pi\pi)/[\Gamma(f_0\to\pi\pi)+\Gamma(f_0\to K\ov K)]=0.68$
\cite{Oller}. Therefore,
 \be
 \B(f_0(980)\to K^+K^-)=0.16\,, \qquad\qquad \B(f_0(980)\to\pi^+\pi^-)=0.45\,.
 \en
For $D\to a_0P$, we apply the PDG (Particle Data Group) average,
$\Gamma(a_0\to K\ov K)/\Gamma(a_0\to \pi\eta)=0.177\pm 0.024$
\cite{PDG}, to obtain
 \be \label{aKK}
 \B(a_0^+(980)\to K^+\ov K^0) &=& \B(a_0^-(980)\to K^- K^0)=0.15\pm 0.02\,,
 \non \\  \B(a_0^0(980)\to K^+K^-) &=& 0.075\pm 0.010\,.
 \en
Needless to say, it is of great importance to have more precise
measurements of the branching fractions of $f_0$ and $a_0$.

{\squeezetable
\begin{table}[pth]
\caption{Experimental branching ratios of various $D\to SP$ decays
measured by ARGUS, E687, E691, E791, CLEO, FOCUS and BaBar, where
use of Eqs. (2.1) and (2.2) for the branching fractions of
$f_0(980)$ and $a_0(980)$ has been made. For simplicity and
convenience, we have dropped the mass identification for
$f_0(980)$, $a_0(980)$ and $K^*_0(1430)$.
 }
\begin{center}
\begin{tabular}{l l l l   }
Collaboration & $\B(D\to SP)\times \B(S\to P_1P_2)$  & $\B(D\to SP)$ \\
 \hline
 E791 & $\B(D^+\to f_0\pi^+)\B(f_0\to\pi^+\pi^-)=(1.9\pm 0.5)\times 10^{-4}$ &
 $\B(D^+\to f_0\pi^+)=(4.3\pm 1.1)\times 10^{-4}$ \\
 FOCUS & $\B(D^+\to f_0K^+)\B(f_0\to K^+K^-)=(3.84\pm 0.92)\times 10^{-5}$ &
 $\B(D^+\to f_0K^+)=(2.4\pm 0.6)\times
 10^{-4}$ \\
 FOCUS & $\B(D^+\to f_0K^+)\B(f_0\to \pi^+\pi^-)=(6.12\pm 3.65)\times 10^{-5}$ &
 $\B(D^+\to f_0K^+)=(1.4\pm 0.8)\times 10^{-4}$ \\
 FOCUS & $\B(D^+\to a_0^0\pi^+)\B(a_0^0\to
 K^+K^-)=(2.38\pm0.47)\times 10^{-3}$ & $\B(D^+\to
 a_0^0\pi^+)=(3.2\pm 0.6)\%$ \\
 E791 & $\B(D^+\to\sigma\pi^+)\B(\sigma\to\pi^+\pi^-)=(1.4\pm0.3)\times 10^{-3}$ &
 $\B(D^+\to\sigma\pi^+)=(2.1\pm0.5)\times 10^{-3}$ \\
 E791 & $\B(D^+\to\kappa\pi^+)\B(\kappa\to K^-\pi^+)=(4.4\pm1.2)\%$ &
 $\B(D^+\to\kappa\pi^+)=(6.5\pm1.9)\%$ \\
 E691,E687 & $\B(D^+\to\ov K_0^{*0}\pi^+)\B(\ov K_0^{*0}\to K^-\pi^+)=(2.3\pm 0.3)\%$  &
 $\B(D^+\to\ov K_0^{*0}\pi^+)=(3.7\pm0.4)\%$ \\
 E791 & $\B(D^+\to\ov K_0^{*0}\pi^+)\B(\ov K_0^{*0}\to K^-\pi^+)=(1.14\pm 0.16)\%$  &
 $\B(D^+\to\ov K_0^{*0}\pi^+)=(1.8\pm0.3)\%$ \\
  \hline
 ARGUS,E687 & $\B(D^0\to f_0\ov K^0)\B(f_0\to \pi^+\pi^-)=(3.2\pm 0.9)\times 10^{-3}$  &
 $\B(D^0\to f_0\ov K^0)=(7.2\pm  2.0)\times 10^{-3}$ \\
 CLEO & $\B(D^0\to f_0\ov K^0)\B(f_0\to \pi^+\pi^-)=(2.5^{+0.8}_{-0.5})\times 10^{-3}$  &
 $\B(D^0\to f_0\ov K^0)=(5.7^{+1.8}_{-1.1})\times 10^{-3}$ \\
 BaBar & $\B(D^0\to f_0\ov K^0)\B(f_0\to K^+K^-)=(1.2\pm 0.9)\times 10^{-3}$  &
 $\B(D^0\to f_0\ov K^0)=(7.4\pm  5.5)\times 10^{-3}$ \\
 BaBar & $\B(D^0\to a_0^+K^-)\B(a_0^+\to K^+\ov K^0)=(3.3\pm 0.8)\times 10^{-3}$ &
 $\B(D^0\to a_0^+K^-)=(2.2\pm0.5)\%$ \\
 BaBar & $\B(D^0\to a_0^-K^+)\B(a_0^-\to K^-\ov K^0)=(3.1\pm1.9)\times 10^{-4}$ &
 $\B(D^0\to a_0^-K^+)=(2.1\pm1.3)\times 10^{-3}$ \\
 BaBar & $\B(D^0\to a_0^0\ov K^0)\B(a_0^0\to K^+K^-)=(5.9\pm 1.3)\times 10^{-3}$ &
 $\B(D^0\to a_0^0\ov K^0)=(7.9\pm1.7)\%$ \\
 BaBar & $\B(D^0\to a_0^+\pi^-)\B(a_0^+\to K^+\ov K^0)=(5.1\pm4.2)\times 10^{-4}$ &
 $\B(D^0\to a_0^+\pi^-)=(3.4\pm2.8)\times 10^{-3}$  \\
 BaBar & $\B(D^0\to a_0^-\pi^+)\B(a_0^-\to K^-K^0)=(1.43\pm1.19)\times 10^{-4}$ &
 $\B(D^0\to a_0^-\pi^+)=(9.5\pm 7.9)\times 10^{-4}$ \\
 ARGUS,E687 & $\B(D^0\to K_0^{*-}\pi^+)\B(K_0^{*-}\to \ov K^0\pi^-)=(7.3\pm 1.6)\times 10^{-3}$ &
 $\B(D^0\to K_0^{*-}\pi^+)=(1.18\pm0.25)\%$ \\
 CLEO & $\B(D^0\to K_0^{*-}\pi^+)\B(K_0^{*-}\to \ov K^0\pi^-)=(4.3^{+1.9}_{-0.8})\times 10^{-3}$ &
 $\B(D^0\to K_0^{*-}\pi^+)=(7.0^{+3.1}_{-1.3})\times 10^{-3}$ \\
 CLEO & $\B(D^0\to K_0^{*-}\pi^+)\B(K_0^{*-}\to K^-\pi^0)=(3.6\pm 0.8)\times 10^{-3}$ &
 $\B(D^0\to K_0^{*-}\pi^+)=(1.17\pm0.26)\%$ \\
 CLEO & $\B(D^0\to \ov K_0^{*0}\pi^0)\B(\ov K_0^{*0}\to K^-\pi^+)=
 (5.3^{+4.2}_{-1.4})\times 10^{-3}$ &
 $\B(D^0\to \ov K_0^{*0}\pi^0)=(8.6^{+6.8}_{-2.3})\times 10^{-3}$  \\
  \hline
 E687 & $\B(D_s^+\to f_0\pi^+)\B(f_0\to K^+K^-)=(4.9\pm 2.3)\times 10^{-3}$ &
 $\B(D_s^+\to f_0\pi^+)=(3.1\pm 1.4)\%$ \\
 E791 & $\B(D_s^+\to f_0\pi^+)\B(f_0\to \pi^+\pi^-)=(5.7\pm 1.7)\times 10^{-3}$ &
 $\B(D_s^+\to f_0\pi^+)=(1.3\pm 0.4)\%$ \\
 FOCUS & $\B(D_s^+\to f_0\pi^+)\B(f_0\to \pi^+\pi^-)=(9.5\pm 2.7)\times 10^{-3}$ &
 $\B(D_s^+\to f_0\pi^+)=(2.1\pm 0.6)\%$ \\
 FOCUS & $\B(D_s^+\to f_0\pi^+)\B(f_0\to K^+K^-)=(7.0\pm1.9)\times 10^{-3}$ &
 $\B(D_s^+\to f_0\pi^+)=(4.4\pm 1.2)\%$ \\
 FOCUS & $\B(D_s^+\to f_0K^+)\B(f_0\to K^+K^-)=(2.8\pm1.3)\times 10^{-4}$ &
 $\B(D_s^+\to f_0K^+)=(1.8\pm0.8)\times 10^{-3}$ \\
 E687 & $\B(D_s^+\to \ov K_0^{*0}K^+)\B(\ov K_0^{*0}\to K^-\pi^+)=(4.3\pm2.5)\times 10^{-3}$ &
 $\B(D_s^+\to \ov K_0^{*0}K^+)=(7\pm4)\times 10^{-3}$  \\
 FOCUS & $\B(D_s^+\to K_0^{*0}\pi^+)\B(K_0^{*0}\to K^+\pi^-)=(1.4\pm 0.8)\times 10^{-3}$ &
 $\B(D_s^+\to K_0^{*0}\pi^+)=(2.3\pm1.3)\times 10^{-3}$  \\
\end{tabular}
\end{center}
\end{table}
}

Several remarks are in order.
 \begin{enumerate}
 \item The Cabibbo doubly suppressed mode $D^+\to K^+K^+K^-$ has been
recently observed by FOCUS \cite{FOCUS}. The Cabibbo-allowed
$D^0\to a_0^+K^-$ and doubly Cabibbo-suppressed mode $D^0\to
a_0^-K^+$ have been extracted from the three-body decay $D^0\to
K^+K^-\ov K^0$ by BaBar \cite{BaBar}.

\item The decay $D^+\to K^-\pi^+\pi^+$ has been measured by E691
\cite{E691} and E687 \cite{E687} and the combined branching ratio
for $D^+\to K^{*0}_0\pi^+$ is quoted to be $(3.7\pm 0.4)\%$ by PDG
\cite{PDG} (see also Table I). A highly unusual feature is that
this three-body decay is dominated by the nonresonant contribution
at 90\% level, whereas it is known that nonresonant effects
account for at most 10\% in other three-body decay modes of
charmed mesons \cite{PDG,a1a2charm}. A recent Dalitz plot analysis
by E791 \cite{E791} reveals that a best fit to the data is
obtained if the presence of an additional scalar resonance called
$\kappa$ is included. As a consequence, the nonresonant decay
fraction drops from 90\% to $(13\pm 6)\%$, whereas $\kappa\pi^+$
accounts for $(48\pm12)\%$ of the total rate. Therefore, the
branching ratio of $D^+\to \ov K^{*0}_0\pi^+$ is dropped from
$(3.7\pm0.4)\%$ to $(1.8\pm0.3)\%$. We shall see in Sec. V that
the form factor for $D\to K^*_0$ transition extracted from the
E791 experiment is more close to the theoretical expectation than
that inferred from E691 and E687.

\item  The new CLEO and BaBar results on the Cabibbo-allowed decay
$D^0\to f_0\ov K^0$ are consistent with the early measurements by
ARGUS \cite{ARGUS} and E687 \cite{E687} quoted in Table I from
PDG. The Cabibbo doubly suppressed mode $D^+\to f_0K^+$ was first
measured by FOCUS recently.

\item There are four measurements of $D^0\to K^{*-}_0\pi^+$: three
from $D^0\to\ov K^0\pi^+\pi^-$ by ARGUS \cite{ARGUS}, E687
\cite{E687}, CLEO \cite{CLEO}, and one from $D^0\to K^-\pi^+\pi^0$
by CLEO. The CLEO result $(1.17\pm 0.26)\%$ for the branching
ratio of $D^0\to K^{*-}_0\pi^+$ extracted from $D^0\to
K^-\pi^+\pi^0$ is in good agreement with ARGUS and E687 (see Table
I), while the CLEO number $(7.0^{+3.1}_{-1.3})\times 10^{-3}$
determined from $D^0\to\ov K^0\pi^+\pi^-$ is slightly lower.

\item As for $D_s^+\to f_0\pi^+$, four measured results by E687,
E791 and FOCUS are shown in Table I. The old measurement by E687
and two new ones by FOCUS are larger than the E791 one. The
preliminary FOCUS measurement indicates that the $f_0(980)$
resonance accounts for $(94.4\pm3.8)\%$ of the total
$D_s^+\to\pi^+\pi^+\pi^-$ rate \cite{FOCUS}. Later we shall use
the average value $\B(D_s^+\to f_0\pi^+)=(1.8\pm 0.3)\%$ in Table
IV.

\item As stressed in the Introduction, there exist three-body
decay modes that are dominated by the scalar resonances. Apart
from the decays $D_s^+\to f_0\pi^+$ and $D^+\to\sigma\pi^+$ as
mentioned in the Introduction, some other examples are $D^+_s\to
f_0(980)K^+$ and $D^+\to f_0(980)K^+$ which account for 72\% and
44.5\%, respectively, of the decays $D_s^+\to K^+K^+K^-$ and
$D^+\to K^+K^+K^-$ \cite{FOCUS}.

\item The production of the resonance $f_0(1370)$ in $D^0\to \ov
K^0\pi^+\pi^-\to f_0(1370)\ov K^0$, $D^+\to K^+K^-\pi^+\to
f_0(1370)\pi^+$ and $D_s^+\to\pi^+\pi^+\pi^-\to f_0(1370)\pi^+$
has been measured by ARGUS, E687, CLEO, by FOCUS and by E791,
respectively. Since the branching fractions of $f_0(1370)$ into
$\pi^+\pi^-,~K^+K^-$ are unknown, we will not discuss it until
Sec. V.D.

\item Some preliminary measurements of $D\to SP$ do not have yet
enough statistical significance, for example, the decays $D^0\to
a_0^\pm \pi^\mp,~a_0^-K^+$ and $D_s^+\to\ov K^{*0}_0K^+$.

 \end{enumerate}

\section{Physical Properties of scalar mesons}

The masses and widths of the $0^+$ scalar mesons relevant for our
purposes are summarized in Table II. The $\sigma$ meson observed
in $D^+\to\pi^+\pi^+\pi^-$ decay by E791 \cite{E791} has a mass of
$478^{+24}_{-23}\pm17$ MeV and a width of $324^{+42}_{-40}\pm21$
MeV. Recently, the decay $D^0\to K_S^0\pi^+\pi^-$ has been
analyzed by CLEO \cite{CLEO}. By replacing the nonresonant
contribution with a $K_S^0\sigma$ component, it is found that
$m_\sigma=513\pm32$ MeV and $\Gamma_\sigma=335\pm67$ MeV
\cite{CLEO}, in accordance with E791. The isodoublet scalar
resonance $\kappa$ observed in the decay $D^+\to K^-\pi^+\pi^+$ by
E791 has a mass of $797\pm19\pm43$ MeV and a width of
$410\pm43\pm87$ MeV \cite{E791}. However, the signal of $\kappa$
is much less evident than $\sigma$. Indeed, this resonance is not
confirmed by CLEO in the Dalitz analysis of the decay $D^0\to
K^-\pi^+\pi^0$ \cite{CLEO}. The well established scalars
$f_0(980)$ and $a_0(980)$ are narrow, while $\sigma$ and $\kappa$
are very broad.

\begin{table}[h]
\caption{The masses and widths of the $1\,^3P_0$ scalar mesons
(except for $\kappa$, see the mini-review in [18]) quoted in [14].
 }
\begin{center}
\begin{tabular}{l c c c c c  }
 & $\sigma$ & $\kappa$ & $f_0(980)$ & $a_0(980)$ &
 $K^*_0(1430)$ \\ \hline
 mass & $400-1200$ MeV & $700-900$ MeV & $980\pm 10$ MeV &
 $984.7\pm 1.2$ MeV & $1412\pm 6$ MeV \\
 width & $600-1000$ MeV & $400-600$ MeV & $40-100$ MeV &
 $50-100$ MeV & $294\pm 23$ MeV \\
\end{tabular}
\end{center}
\end{table}

\subsection{Quark structure of scalar mesons}
It is known that the identification of scalar mesons is difficult
experimentally and the underlying structure of scalar mesons is
not well established theoretically (for a review, see e.g.
\cite{Spanier,Godfrey,Close}). It has been suggested that the
light scalars--the isoscalars $\sigma(500)$, $f_0(980)$, the
isodoublet $\kappa$ and the isovector $a_0(980)$--form an SU(3)
flavor nonet. In the naive quark model, the flavor wave functions
of these scalars read
 \be
 && \sigma={1\over \sqrt{2}}(u\bar u+d\bar d), \qquad\qquad f_0= s\bar s\,, \non \\
 && a_0^0={1\over\sqrt{2}}(u\bar u-d\bar d), \qquad a_0^+=u\bar d, \qquad
 a_0^-=d\bar u,  \\
 && \kappa^{+}=u\bar s, \qquad \kappa^{0}= d\bar s, \qquad
 \bar \kappa^{0}=s\bar d,\qquad \kappa^{-}=s\bar u. \non
 \en
However, this model immediately faces two
difficulties:\footnote{However, for a different point of view of
these difficulties in the $q\bar q$ picture, see e.g.
\cite{Beveren1}.} (i) It is impossible to understand the mass
degeneracy of $f_0(980)$ and $a_0(980)$. (ii) It is hard to
explain why $\sigma$ and $\kappa$ are broader than $f_0(980)$ and
$a_0(980)$. Recalling that $a_0\to\pi\eta,~\sigma\to\pi\pi$ and
$\kappa\to K\pi$ are OZI allowed (but not OZI superallowed !)
while $f_0\to\pi\pi$ is OZI suppressed as it is mediated by the
exchange of two gluons, it is thus expected that $m_\kappa\gg
\Gamma_\kappa\sim\Gamma_\sigma\sim\Gamma_{a_0}>\Gamma_{f_0}$, a
relation not borne out by experiment.

Although the data of $D_s^+\to f_0(980)\pi^+$ and $\phi\to
f_0(980)\gamma$ imply the copious $f_0(980)$ production via its
$s\bar s$ component, there are some experimental evidences
indicating that $f_0(980)$ is not purely an $s\bar s$ state.
First, the measurements of $J/\psi\to f_0(980)\phi$ and $J/\psi\to
f_0(980)\omega$
 \be \label{Jpsi}
 \B(J/\psi\to f_0(980)\phi) &=& (3.2\pm0.9)\times 10^{-4}, \non \\
 \B(J/\psi\to f_0(980)\omega) &=& (1.4\pm0.5)\times 10^{-4}
 \en
clearly indicate the existence of the non-strange and strange
quark content in $f_0(980)$. Second, the fact that $f_0(980)$ and
$a_0(980)$ have similar widths and that the $f_0$ width is
dominated by $\pi\pi$ also suggests the composition of $u\bar u$
and $d\bar d$ pairs in $f_0(980)$; that is, $f_0(980)\to\pi\pi$
should not be OZI suppressed relative to $a_0(980)\to\pi\eta$.
Therefore, isoscalars $\sigma$ and $f_0$ must have a mixing
 \be \label{f0sigmaw.f.}
 f_0 = s\bar s\,\cos\theta+n\bar n\sin\theta,\qquad
 \sigma &=& -s\bar s\,\sin\theta+n\bar n\cos\theta,
 \en
with $n\bar n\equiv (u\bar u+d\bar d)/\sqrt{2}$.

The $\sigma\!-\!f_0(980)$ mixing angle can be inferred from the
decays $J/\psi\to f_0(980)\phi$ and $J/\psi\to f_0(980)\omega$
\cite{Achasov} \footnote{It has been shown by the $f_0(980)$
production data in $Z^0$ decays at OPAL \cite{OPAL} and DELPHI
\cite{DELPHI} that $f_0(980)$ is composed essentially of $u\bar u$
and $d\bar d$ pairs. This favors a mixing angle close to $\pi/2$.
However, it is in contradiction to the experimental observation
that the final state $f_0(980)\phi$ in hadronic $J/\psi$ decays
has a larger rate than $f_0(980)\omega$.}
 \be
 {\B(J/\psi\to f_0(980)\omega)\over \B(J/\psi\to
 f_0(980)\phi)}={1\over \lambda}\tan^2\theta,
 \en
where the deviation of the parameter $\lambda$ from unity
characterizes the suppression of the $s\bar s$ pair production;
that is, $\lambda=1$ in the SU(3) limit. From the data
(\ref{Jpsi}) we obtain
 \be \label{f0theta1}
 \theta=(34\pm 6)^\circ, \qquad {\rm or}\quad
 \theta=(146\pm6)^\circ
 \en
for $\lambda=1$. Another information on the mixing angle can be
obtained from the $f_0(980)$ coupling to $\pi\pi$ and $K\ov K$
\cite{Achasov}:
 \be
 R_g\equiv{g_{f_0K^+K^-}^2\over
 g_{f_0\pi^+\pi^-}^2}={1\over 4}(\lambda+\sqrt{2}\cot\theta)^2.
 \en
Using the average value of $R_g=4.03\pm0.14$ obtained from the
measurements: $4.00\pm0.14$ by KLOE \cite{KLOE1}, $4.6\pm0.8$ by
SND \cite{SND} and $6.1\pm 2.0$ from CMD-2 \cite{CMD}, we find
 \be \label{f0theta2}
 \theta=(25.1\pm0.5)^\circ, \qquad {\rm or} \quad
 \theta=(164.3\pm0.2)^\circ
 \en
for $\lambda=1$. However, the WA102 experiment on $f_0(980)$
production in central $pp$ collisions yields a result
$R_g=1.63\pm0.46$ \cite{WA102}, which differs from the
aforementioned three measurements. This leads to
 \be \label{f0theta3}
  \theta=(42.3^{+8.3}_{-5.5})^\circ, \qquad {\rm or} \quad
 \theta=(158\pm2)^\circ,
 \en
again for $\lambda=1$.

Recently, a phenomenological analysis of the radiative decays
$\phi\to f_0(980)\gamma$ and $f_0(980)\to\gamma\gamma$ yields
 \be \label{f0theta4}
 \theta=(5\pm 5)^\circ,\qquad {\rm or} \quad\theta=(138\pm6)^\circ
 \en
with the second solution being more preferable \cite{Anisovich}.
In this analysis, $\phi\to f_0(980)\gamma$ is calculated at the
quark level by considering the $s\bar s$ quark loop coupled to
both $\phi$ and $f_0(980)$.\footnote{It is pointed out in
\cite{Achasov01} that the mechanism $\phi\approx s\bar s\to s\bar
s\gamma\to f_0(980)\gamma$ without creation of and annihilation of
an additional $u\bar u$ pair cannot explain the $f_0(980)$
spectrum observed in $\phi\to \gamma f_0(980)\to\gamma\pi^0\pi^0$
process because it does not contain the $K^+K^-$ intermediate
state. For a criticism of \cite{Anisovich}, see also the remark in
the footnote [28] in the first paper of \cite{Achasov01}.}
However, the experimental analysis and the theoretical study of
this $\phi$ radiative decay are practically based on the
chiral-loop picture, namely, $\phi\to K^+K^-\to K^+K^-\gamma\to
f_0(980)\gamma$. It turns out that the predicted branching ratio
in the $q\bar q$ picture is at most of order $5\times 10^{-5}$
(see e.g. \cite{Achasov97}), while experimentally \cite{PDG}
 \be \label{phif0gamma}
 \B(\phi\to f_0(980)\gamma)=(3.3^{+0.8}_{-0.5})\times 10^{-4}.
 \en
This is because the $f_0(980)$ coupling to $K^+K^-$ is not strong
enough as in the case of the four-quark model to be discussed
shortly where $f_0\to K^+K^-$ is OZI superallowed.

In short, it is not clear if there exists a universal mixing angle
$\theta$ which fits simultaneously to all the measurements from
hadronic $J/\psi$ decays, the $f_0(980)$ coupling to $\pi^+\pi^-$
and $K^+K^-$ and the radiative decay $\phi\to f_0(980)\gamma$
followed by $f_0\to\gamma\gamma$.\footnote{The analysis of the
three-body decays of $D^+\to f_0(980)\pi^+,~\ov
K^{*0}_0(1430)\pi^+$ and $D_s^+\to f_0(980)\pi^+$ gives
$\theta=(42.14^{+5.8}_{-7.3})^\circ$ in \cite{Minkowski}. However,
in this analysis, the weak annihilation contribution has been
neglected and SU(3) symmetry has been applied to relate $f_0\pi$
to $K^*_0\pi$, a procedure which is not justified since $f_0$ and
$K_0^*$ belong to different SU(3) flavor nonets (see the
discussion in Sec. IV). If the mass parameters $m_{n\bar n}$ and
$m_{s\bar s}$ are assigned to the $n\bar n$ and $s\bar s$ states,
respectively, one will have the mass relations: $m_{n\bar
n}^2=m_\sigma^2\cos\theta^2+m_{f_0}^2\sin\theta^2$ and $m_{s\bar
s}^2=m_\sigma^2\sin\theta^2+m_{f_0}^2\cos\theta^2$ from Eq.
(\ref{f0sigmaw.f.}). Inserting the dynamically generated NJL-type
masses for $m_{n\bar n}$ and $m_{s\bar s}$, it is found in
\cite{Scadron} that $\theta=\pm(18\pm 2)^\circ$ provided that
$m_\sigma=600$ MeV. } Eqs. (\ref{f0theta1}) and (\ref{f0theta4})
indicate that $\theta\sim 140^\circ$ is preferred, while
$\theta\sim 34^\circ$ is also allowed provided that $R_g$ is of
order 2. At any rate, the above two possible allowed angles imply
the dominance of $s\bar s$ in the $f_0(980)$ wave function and
$n\bar n$ in $\sigma$. However, as stressed before, the 2-quark
picture for $f_0(980)$ has the difficulty of explaining the
absolute $\phi\to f_0(980)\gamma$ rate.

As for $a_0(980)$, it appears at first sight that one needs an
$s\bar s$ content in $a_0$ in order to explain the radiative decay
$\phi\to a_0(980)\gamma$; otherwise, it is OZI suppressed.
However, since $a_0$ is an isovector while $s\bar s$ is isoscalar,
the mixing of $(u\bar u-d\bar d)$ with $s\bar s$ is not allowed in
the $a_0$ wave function within the 2-quark
description.\footnote{Even in the presence of the hidden $s\bar s$
content in $a_0(980)$ within the 4-quark model, the direct
radiative decay $\phi\to a_0(980)\gamma$ is prohibited owing to
the opposite sign between the $u\bar u$ and $d\bar d$ components
in $a_0(980)$. This means that it is necessary to consider the
contribution from the $K^+K^-$ intermediate states.} Nevertheless,
$\phi\to a_0(980)\gamma$ can proceed through the process $\phi\to
K^+K^-\to K^+K^-\gamma\to a_0(980)\gamma$ as both $\phi$ and
$a_0(980)$ couple to $K^+K^-$. Indeed, it has been suggested that
both $a_0(980)$ and $f_0(980)$ can be interpreted as a $K\ov K$
molecular bound state which is treated as an extended object.
Since both $f_0(980)$ and $a_0(980)$ couple strongly to $K\ov K$
as they are just below the $K\ov K$ threshold, they can be
imagined as an $s\bar s$ and $n\bar n$ core states, respectively,
surrounded by a virtual $K\ov K$ cloud \cite{Weinstein}. In this
$K\ov K$ molecular picture, one can explain the decay $f_0\to
\pi\pi$ without the light non-strange quark content in $f_0(980)$
and the decay $\phi\to a_0(980)\gamma$ without the need of an
intrinsic strange quark component in $a_0$; both decays are
allowed by the OZI rule in the sense that only one gluon exchange
is needed.

However, there are several difficulties with this $K\ov K$
molecular picture. First, the $K\ov K$ molecular width is less
than its binding energy of order 20 MeV \cite{Weinstein}, while
the measured widths of $f_0(980)$ and $a_0(980)$ lie in the range
of 40 to 100 MeV \cite{PDG}. Second, it is expected in this model
that $\B(\phi\to f_0(980)\gamma)/\B(\phi\to a_0(980)\gamma)\approx
1$, while this ratio is measured to be $3.8\pm1.0$ \cite{PDG}.
(The most recent result is $6.1\pm 0.6$ by KLOE \cite{KLOE}.)
Third, the predicted branching ratios for both $\phi\to
f_0(980)\gamma$ and $\phi\to a_0(980)\gamma$ are only of order
$10^{-5}$ \cite{Achasov97} which are too small compared to
(\ref{phif0gamma}) and \cite{Achasov00}
 \be \label{phia0gamma}
 \B(\phi\to a_0(980)\gamma)=(0.88\pm
0.17)\times 10^{-4}.
 \en

Alternatively, the aforementioned difficulties\footnote{Likewise,
it has been argued in the literature that $\sigma$ is not a $q\bar
q$ state \cite{Shakin}. Furthermore, the QCD sum rule calculation
also indicates that the lightest scalars are nearly decoupled from
$q\bar q$, suggesting a non-$q\bar q$ structure \cite{Shi}. In
short, one always has some troubles when the light scalar mesons
are identified as $q\bar q$ states.} with $a_0$ and $f_0$ can be
circumvented in the four-quark model in which one writes
symbolically \cite{Jaffe}
 \be \label{4quarkw.f.}
 && \sigma=ud\bar u\bar d, \qquad\qquad f_0=(us\bar u\bar s+ds\bar d\bar s)/\sqrt{2},  \non \\
 && a_0^0=(us\bar u\bar s-ds\bar d\bar s)/\sqrt{2}, \qquad a_0^+=us\bar d\bar s,
 \qquad a_0^-=ds\bar u\bar s, \non \\
 && \kappa^+=ud\bar d\bar s, \qquad \kappa^0=ud\bar u\bar s,
 \qquad \bar \kappa^0=us\bar u\bar d, \qquad \kappa^-=ds\bar u\bar
 d.
 \en
This is supported by a recent lattice calculation \cite{Alford}.
This $q^2\bar q^2$ scenario for light scalars has several major
advantages: (i) The mass degeneracy of $f_0(980)$ and $a_0(980)$
is natural and the mass hierarchy pattern of the SU(3) nonet is
understandable. (ii) Why $\sigma$ and $\kappa$ are broader than
$f_0$ and $a_0$ can be explained. The decays $\sigma\to\pi\pi$,
$\kappa\to K\pi$ and $f_0,a_0\to K\ov K$ are OZI superallowed
without the need of any gluon exchange, while $f_0\to\pi\pi$ and
$a_0\to\pi\eta$ are OZI allowed as it is mediated by one gluon
exchange. Since $f_0(980)$ and $a_0(980)$ are very close to the
$K\ov K$ threshold, the $f_0(980)$ width is dominated by the
$\pi\pi$ state and $a_0$ governed by the $\pi\eta$ state.
Consequently, their widths are narrower than $\sigma$ and
$\kappa$. (iii) It predicts the relation
 \be
 {\B(J/\psi \to f_0(980)\omega)\over \B(J/\psi\to
 f_0(980)\phi)}={1\over 2},
 \en
which is in good agreement with the experimental value of $0.44\pm
0.20$ \cite{PDG}. (iv) The coupling of $f_0$ and $a_0$ to $K\ov K$
is strong enough as the strong decays $f_0\to K\ov K$ and $a_0\to
K\ov K$ are OZI supperallowed. Consequently, the branching ratio
of $\phi\to (f_0,a_0)\gamma$ can be as large as of order $10^{-4}$
\cite{Achasov97}.\footnote{Just as in the $K\ov K$ molecular
model, the ratio  $r\equiv \B(\phi\to f_0\gamma)/\B(\phi\to
a_0\gamma)$ is also an issue in the four-quark model in which
$g_{f_0K^+K^-}=g_{a_0 K^+K^-}$ and hence $\B(\phi\to
f_0\gamma)=\B(\phi\to a_0\gamma)$ is predicted, in disagreement
with the observed value of $3.8\pm1.0$ \cite{PDG}. Close and Kirk
\cite{Kirk} proposed that $r$ can be explained by considering a
large $a_0\!-\!f_0$ mixing. However, as pointed out in
\cite{Achasov02}, the isospin-violating $a_0\!-\!f_0$ mixing is
small, analogous to the smallness of $\pi^0\!-\!\eta\!-\!\eta'$
mixing, and its correction to $r$ amounts to at most a few
percent. One possibility for a large $r$ is that the superposition
of the 4-quark state and $K\ov K$ has a different weight in
$f_0(980)$ and $a_0(980)$.} (v) It is concluded in
\cite{Achasov01} that production of $f_0(980)$ and $a_0(980)$ in
the $\phi\to\gamma f_0(980)\to\gamma\pi^0\pi^0$ and $\phi\to\gamma
a_0(980)\to\gamma\pi^0\eta$ decays is caused by the four-quark
transitions, resulting in strong restrictions on the large-$N_c$
expansions of the decay amplitudes. The analysis shows that these
constraints give new evidences in favor of the four-quark picture
of $f_0(980)$ and $a_0(980)$ mesons.

Therefore, it appears that the four-quark state in core with the
$K\ov K$ in the outer regime gives a more realistic description of
the light scalar mesons. If scalar mesons near and below 1 GeV are
non-$q\bar q$ states, then the $0^+$ mesons in the $1.3-1.7$ GeV
mass region may be more conventional. For example, it is natural
to assume that $f_0(1370)$, $a_0(1450)$, $K^*_0(1430)$ and
$f_0(1500)/f_0(1710)$ are in the same SU(3) flavor nonet in the
states $n\bar n$, $u\bar d$, $u\bar s$ and $s\bar s$, respectively
\cite{Spanier}. In other words, they may have a simple $q\bar q$
interpretation. A global picture emerged from above discussions is
as follows: The scalar meson states above 1 GeV form a $q\bar q$
nonet with some possible mixing with glueballs, whereas the light
scalar mesons below or near 1 GeV form predominately a $qq\bar
q\bar q$ nonet with a possible mixing with $0^+$ $q\bar q$ and
glueball states (see also \cite{Close}). This is understandable
because in the $q\bar q$ quark model, the $0^+$ meson has a unit
of orbital angular momentum and hence it should have a higher mass
above 1 GeV. On the contrary, four quarks $q^2\bar q^2$ can form a
$0^+$ meson without introducing a unit of orbital angular
momentum. Moreover, color and spin dependent interactions favor a
flavor nonet configuration with attraction between the $qq$ and
$\bar q\bar q$ pairs. Therefore, the $0^+$ $qq\bar q\bar q$ nonet
has a mass near or below 1 GeV.

It is conceivable that the two-quark and four-quark descriptions
of light scalars, especially $f_0(980)$ and $a_0(980)$, may lead
to some different implications for the hadronic weak decays of
charmed mesons into the final state containing a scalar meson.
This will be explored in Sec. V.

\subsection{Decay constants}
The scalar mesons under consideration are $\sigma(500)$, $\kappa$,
$f_0(980)$, $a_0(980)$ and $K^*_0(1430)$. The decay constants of
scalar and pseudoscalar mesons are defined by
 \be \label{decaycon}
  \la 0|A_\mu|P(q)\ra &=& if_Pq_\mu, \qquad\qquad\quad \la
0|V_\mu|S(q)\ra= f_S q_\mu.
 \en
For the neutral scalars $\sigma$, $f_0$ and $a_0^0$, the decay
constant must be zero owing to charge conjugation invariance or
conservation of vector current:
 \be
 f_{\sigma}=f_{f_0}=f_{a_0^0}=0.
 \en
Applying the equation of motion, it is easily seen that the decay
constant of $K^{*+}_0$ ($a^+_0$) is proportional to the mass
difference between the constituent $s$ ($d$) and $u$ quarks.
Contrary to the case of pseudoscalar mesons, the decay constant of
the scalar meson vanishes in the SU(3) limit or even in the
isospin limit. Therefore, the decay constant of $K^*_0(1430)$ and
the charged $a_0(980)$ is suppressed. We shall use the values
 \be
 f_{a^\pm_0}=1.1\,{\rm MeV}, \qquad f_{K^*_0}=42\,{\rm MeV}
 \en
obtained from the finite-energy sum rules \cite{Maltman}. (A
different calculation of the scalar meson decay constants based on
the generalized NLJ model is given in \cite{Shakin1}.) Since they
are derived using the $q\bar q$ quark model, it is not clear if
the $a_0^\pm$ decay constant remains the same in the $q^2\bar q^2$
picture, though it is generally expected that the decay constant
is suppressed in the latter scenario because a four-quark state is
larger than a two-quark state \cite{Maltman}.

As for the decay constant of $\kappa$, we apply the equation of
motion to Eq. (\ref{decaycon}) to obtain
 \be
 m_{a_0}^2f_{a_0} = i(m_d-m_u)\la a_0^-|\bar du|0\ra, \qquad\qquad
 m_{\kappa}^2f_{\kappa} = i(m_s-m_u)\la \kappa^-|\bar su|0\ra,
 \en
and assume $\la \kappa^-|\bar s u|0\ra\approx \la a_0^-|\bar d
u|0\ra$. It follows that $f_\kappa\approx 65$ MeV for $m_u=4.8$
MeV, $m_d=8.7$ MeV, $m_s=164$ MeV \cite{Diehl} and $m_\kappa=800$
MeV. In short, the decay constants of scalar mesons are either
zero or very small.

\subsection{Form factors}
Form factors for $D\to P$ and $D\to S$ transitions are defined by
\cite{BSW}
 \be \label{m.e.}
 \la P(p)|V_\mu|D(p_D)\ra &=& \left(p_{D\mu}+p_\mu-{m_D^2-m_{P}^2\over q^2}\,q_ \mu\right)
F_1^{DP}(q^2)+{m_D^2-m_{P}^2\over q^2}q_\mu\,F_0^{DP}(q^2), \non \\
 \la S(p)|A_\mu|D(p_D)\ra &=&
i\left[\left(p_{D\mu}+p_\mu-{m_D^2-m_S^2\over q^2}\,q_ \mu\right)
F_1^{DS}(q^2)+{m_D^2-m_S^2\over q^2}q_\mu\,F_0^{DS}(q^2)\right],
 \en
where $q_\mu=(p_D-p)_\mu$. The form factors relevant for $D\to SP$
decays are $F_0^{DP}(q^2)$ and $F_0^{DS}(q^2)$. For $D\to P$ form
factors, we will use the Melikhov-Stech  (MS) model \cite{MS}
based on the constituent quark picture. Other form factor models
give similar results.

As discussed in Sec. III.A, the light scalar mesons
$\sigma,~\kappa,~f_0(980)$ and $a_0(980)$ are predominately
$q^2\bar q^2$, while $K^*_0$ is described by the $q\bar q$ state.
Nevertheless, it is useful to see what are the predictions of
$D\to S$ form factors in the conventional quark model. There are
some existing calculations of the form factor $F_0^{DS}(0)$ in the
literature (see Table III). Paver and Riazuddin \cite{Paver}
obtained $F_0^{D\sigma}(0)=0.74(f_D/200\,{\rm MeV})$. Gatto {\it
et al.} got $F_0^{D\sigma}(0)=0.57\pm0.09$ \cite{Gatto} using the
constituent quark model (CQM). Based on the same model, Deandrea
{\it et al.} \cite{Deandrea} obtained $F_0^{D_s^+
f_0}(0)=0.64^{+0.05}_{-0.03}$ assuming a pure $s\bar s$ state for
$f_0$. A value of $F_0^{D_s^+ f_0}(m_\pi^2)=0.36^{+0.06}_{-0.08}$
is obtained by Gourdin, Keum and Pham \cite{Gourdin} based on a
fit to the old data of $D_s^+\to f_0(980)\pi^+$. The $B\to a_0$
form factor is estimated by Chernyak \cite{Chernyak} to be
$F_0^{Ba_0}(0)\sim 0.46$. Using the scaling law, it leads to
$F_0^{Da_0}(0)\approx F_0^{Ba_0}(0)\sqrt{m_B/m_D}=0.77$. Since the
conventional quark model is not applicable to light scalars with
four-quark content, we shall use the measured decay rates to
extract the $D\to S$ form factors (except for $D\to a_0$) in Sec.
V and the results are summarized in Table III.

\begin{table}[ht]
\caption{The $D\to S$ transition form factors $F_0^{DS}(0)$ at
$q^2=0$ in various models. Except for the $D\to a_0$ form factor,
the other form factors in this work are obtained by a fit to the
data. The $D\to f_0$ form factor is obtained from the $D_s^+\to
f_0$ one via Eq. (3.20).
 }
\begin{center}
\begin{tabular}{l c c  c  c }
Transition & \cite{Paver} & \cite{Gatto} & \cite{Deandrea} & This work \\
\hline
$D\to\sigma$ & 0.74 & $0.57\pm 0.09$ & & $0.42\pm 0.05$ \\
$D\to f_0$ & & & & $0.26\pm0.02$ \\
$D_s^+\to f_0$ & & & $0.64^{+0.05}_{-0.03}$ & $0.52\pm0.04$ \\
$D\to a_0^\pm$ & & & & 0.77 \\
$D\to \kappa$ & & & & $0.85\pm0.10$ \\
$D,D_s^+\to K_0^*$ & & & & $1.20\pm 0.07$  \\
\end{tabular}
\end{center}
\end{table}

In the $q\bar q$ description of $f_0(980)$, it follows from Eq.
(\ref{f0sigmaw.f.}) that
 \be
 F_0^{D^0f_0}={1\over\sqrt{2}}\sin\theta\,F_0^{D^0f_0^{u\bar u}}, \qquad
 F_0^{D^+f_0}={1\over\sqrt{2}}\sin\theta\,F_0^{D^+f_0^{d\bar d}},
 \qquad
 F_0^{D_s^+f_0}=\cos\theta\,F_0^{D_s^+f_0^{s\bar s}},
 \en
where the superscript $q\bar q$ denotes the quark content of $f_0$
involved in the transition. In the limit of SU(3) symmetry,
$F_0^{D^0f_0^{u\bar u}}=F_0^{D^+f_0^{d\bar
d}}=F_0^{D^+_sf_0^{s\bar s}}$ and hence
 \be
 F_0^{D^0f_0}=F_0^{D^+f_0}={1\over\sqrt{2}}\,F_0^{D_s^+f_0}\,\tan\theta.
 \en
In the four-quark picture, one has (see Fig. 1)
 \be \label{Df0}
 F_0^{Df_0}(0)={\lambda\over 2}F_0^{D_s^+f_0}(0),
 \en
where use of the $f_0(980)$ flavor function $s\bar s(u\bar u+d\bar
d)/\sqrt{2}$ has been made. For $\theta\sim 34^\circ$, we see that
the relation between $F_0^{Df_0}$ and $F_0^{D_s^+f_0}$ is very
similar in the $q\bar q$ and $q^2\bar q^2$ pictures, but if
$\theta\sim 140^\circ$ then $F_0^{Df_0}$ will have an opposite
sign to $F_0^{D_s^+f_0}$ in the former model.

\begin{figure}[t]
\hspace{3cm}
  \psfig{figure=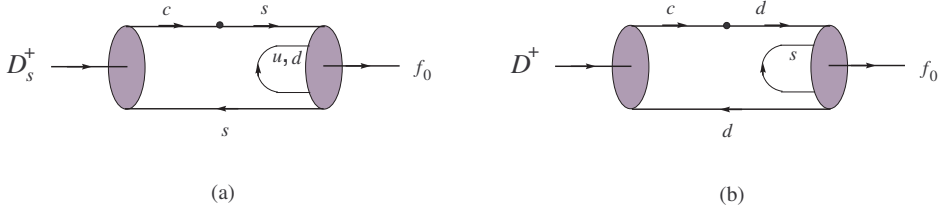,width=6in}
\vspace{-7cm}
    \caption[]{\small The $D_s^+\to f_0(980)$ and $D^+\to
    f_0(980)$ transition form factors, where $f_0(980)$ is
    described by a $q^2\bar q^2$ state.
    }
\end{figure}

To compute the $D\to S$ form factors using the $q\bar q$ model, it
is worth mentioning the Isgur-Scora-Grinstein-Wise (ISGW) model
\cite{ISGW} and its improved version the ISGW2 model \cite{ISGW2}.
Contrary to the above-mentioned form-factor calculations shown in
Table III, the form factors of interest in the ISGW model are
$F_1^{DS}$ evaluated at $q^2=q^2_m\equiv (m_D-m_S)^2$, the maximum
momentum transfer, recalling that $F_1(0)=F_0(0)$. The reason for
considering the form factor at zero recoil is that the form-factor
$q^2$ dependence in the ISGW model is proportional to
Exp[$-(q^2_m-q^2)$]. Hence, the form factor decreases
exponentially as a function of $(q^2_m-q^2)$. Consequently, the
form factor is unreasonably small at $q^2=0$. This has been
improved in the ISGW2 model in which the form factor has a more
realistic behavior at large $(q^2_m-q^2)$. However, we find that
form factors in the ISGW2 model calculated even at zero recoil are
already small compared to the other model calculations at $q^2=0$.
For example, $F_1^{Da_0}$ and $F_1^{D_s K_0^*}$ at zero recoil are
found to be 0.52 and 0.29 respectively in the ISGW model and 0.12
as well as 0.09 respectively in the ISGW2 model. These form
factors become even much smaller at $q^2=0$. Therefore, the ISGW
model and especially the ISGW2 version predict much smaller $D\to
S$ form factors than other quark models.

For the $q^2$ dependence of the $D\to S$ form factors, we shall
assume the pole dominance:
 \be
 F^{DS}_0(q^2)={F^{DS}_0(0)\over 1-q^2/m_*^2}\,,
 \en
with $m_*$ being the mass of the $0^-$ pole state with the same
quark content of the current under consideration.

In the MS \cite{MS} or the BSW \cite{BSW} model, the typical $D\to
P$ form factors have the values $F_0^{D\pi^\pm}(0)=0.69$ and
$F_0^{DK}(0)=0.78$. In general, it is conceivable that the form
factor for $D\to \sigma$ transition is comparable to the $D\to
\pi^0$ one. The argument goes as follows. If the scalar meson is
made from $q\bar q$, its distribution amplitude has the form
\cite{Diehl}
 \be
 \phi_S(x)=6f_S\, x(1-x)\left[1+\sum_{n=1}^\infty
 B_n\,C_n^{3/2}(2x-1)\right],
 \en
where $f_S$ is the decay constant of the scalar meson $S$, $B_n$
are constants, and $C_n^{3/2}$ is the Gegenbauer polynomial. For
the isosinglet scalar mesons $\sigma$ and $f_0$, their decay
constants vanish, but the combination $f_SB_n$ can be nonzero. For
$\sigma$, $f_0$ and $a_0^0$, charge conjugation invariance implies
that $\phi(x)=-\phi(1-x)$; that is, the distribution amplitude
vanishes at $x=1/2$. Using $|f_SB_1|\approx$ 75 MeV obtained in
\cite{Diehl}, we have
 \be
 \phi_\sigma(x)\approx 6\tilde B_1 x(1-x)(3-6x),
 \en
where $\tilde B_1=-f_SB_1$. It is clear that the $\sigma$
distribution amplitude peaks at $x=0.25$ and $0.75$. Now the $D$
meson wave function is peaked at $x\sim \Lambda/m_D\sim 1/3$
\cite{HnLi}. Recall that the asymptotic pion distribution
amplitude has the familiar expression
 \be
 \phi_\pi(x)= 6f_\pi x(1-x),
 \en
which has a peak at $x=1/2$. Though $\tilde B_1$ is smaller than
$f_\pi$, it is anticipated that the $D\to\sigma$ transition form
factor is similar to that of $D\to \pi^0$ one because the peak of
$\phi_\sigma$ is close to that of the $D$ distribution amplitude.
However, it is not clear if this argument still holds for the
scalar mesons which are bound states of $q^2\bar q^2$.

For $K^*_0(1430)$, the distribution amplitude reads
 \be
 \phi_{K^*_0}(x)\approx 6f_{K^*_0}x(1-x)\left[1+B_1(3-6x)\right].
 \en
It is easy to check that $\phi_{K^*_0}$ has a large peak at
$x=0.25$ and a small peak at $x=0.75$. Consequently, it is natural
to have a large $D\to K^*_0$ transition form factor as shown in
Table III. Another argument favoring a large $D\to K^*_0$ form
factor is as follows (see \cite{Kamal}). Consider the decay
$D^0\to K^{*-}_0\pi^+$ and apply PCAC to evaluate the matrix
element $q^\mu\la K^*_0|\bar s\gamma_\mu(1-\gamma_5)c|D^0\ra$.
Assuming that this two-body matrix element is saturated by the
$D_s^+$ pole, we find
 \be
 \la \pi^+|(\bar ud)|0\ra\la K^{*-}_0|(\bar
 sc)|D^0\ra=i\sqrt{2}f_\pi f_{D_s}\,g_{K^{*-}_0D^0D_s^+}\,{m^2_{D_s}\over
 m_{D_s}^2-q^2}.
 \en
Next apply the SU(4) symmetry to relate strong coupling of
$K^{*-}_0D^0D_s^+$ to $K^{*-}_0\pi^+\ov K^0$
 \be
 g_{K^{*-}_0D^0D_s^+}=g_{K^{*-}_0\pi^+\ov K^0}.
 \en
The coupling $g_{K^{*-}_0\pi^+\ov K^0}$ can be determined from the
measured decay rate of $K^{*-}_0\to \ov K^0\pi^-$ via
 \be
 \Gamma(K^{*-}_0\to \ov K^0\pi^-)=g_{K^{*-}_0\pi^+\ov
 K^0}^2\,{p_c\over 8\pi m^2_{K^*_0}},
 \en
where $p_c$ is the c.m. momentum in the $K^*_0$ rest frame. Using
the $K^*_0$ width given in Table II, we obtain
$g_{K^{*-}_0\pi^+\ov K^0}=4.9$ GeV. Since
 \be
  \la \pi^+|(\bar ud)|0\ra\la K^{*-}_0|(\bar
 sc)|D^0\ra=f_\pi (m_D^2-m^2_{K^*_0})F_0^{DK^*_0}(m_\pi^2),
 \en
it follows that
 \be
 F_0^{DK^*_0}(m_\pi^2)=\sqrt{2}\,{ g_{K^{*-}_0\pi^+\ov K^0}f_{D_s^+}
 \over m_D^2-m^2_{K^*_0}}\,{m^2_{D_s}\over
 m^2_{D_s}-m_\pi^2}=1.23\left({f_{D^+_s}\over 270\,{\rm
 MeV}}\right).
 \en
This is consistent with the value of $1.20\pm0.07$ extracted
directly from $D^+\to \ov K^{*0}_0\pi^+$ (see Sec. V.A).

\section{Quark diagram scheme and final-state interactions}

It has been established sometime ago that a least
model-independent analysis of heavy meson decays can be carried
out in the so-called quark-diagram approach. In this diagrammatic
scenario, all two-body nonleptonic weak decays of heavy mesons can
be expressed in terms of six distinct quark diagrams
\cite{Chau,CC86,CC87}: $T$, the color-allowed external
$W$-emission tree diagram; $C$, the color-suppressed internal
$W$-emission diagram; $E$, the $W$-exchange diagram; $A$, the
$W$-annihilation diagram; $P$, the horizontal $W$-loop diagram;
and $V$, the vertical $W$-loop diagram. (The one-gluon exchange
approximation of the $P$ graph is the so-called ``penguin
diagram".) It should be stressed that these quark diagrams are
classified according to the topologies of weak interactions with
all strong interaction effects included and hence they are {\it
not} Feynman graphs. All quark graphs used in this approach are
topological and meant to have all the strong interactions
included, i.e. gluon lines are included in all possible ways.
Therefore, topological graphs can provide information on
final-state interactions (FSIs).

Based on the SU(3) flavor symmetry, this model-independent
analysis enables us to extract the topological quark-graph
amplitudes and see the relative importance of different underlying
decay mechanisms. The quark-diagram scheme, in addition to be
helpful in organizing the theoretical calculations, can be used to
analyze the experimental data directly. When enough measurements
are made with sufficient accuracy, we can find out the values of
each quark-diagram amplitude from experiment and compare to
theoretical results, especially checking whether there are any
final-state interactions or whether the weak annihilations can be
ignored as often asserted in the literature.

For charmed meson decays, the penguin contributions are negligible
owing to the good approximation $V_{ud}V_{cd}^*\approx
-V_{us}V_{cs}^*$ and the smallness of $V_{ub}V_{cb}^*$. Hence, for
$D\to PP,VP,VV$ decays, only $T$, $C$, $E$ and $A$ contribute. The
reduced quark-graph amplitudes $T,C,E,A$ have been extracted from
Cabibbo-allowed $D\to PP$ decays by Rosner \cite{Rosner,Rosner02}
with the results:
 \be \label{PP}
 T &=& (2.67\pm0.20)\times 10^{-6}\,{\rm GeV}, \non \\
 C &=& (2.03\pm0.15)\,{\rm Exp}[-i(151\pm4)^\circ]\times
 10^{-6}\,{\rm GeV}, \non \\
 E &=& (1.67\pm0.13)\,{\rm Exp}[\,i(115\pm5)^\circ]\times 10^{-6}\,{\rm GeV},
 \non \\ A &=& (1.05\pm0.52)\,{\rm Exp}[-i(65\pm30)^\circ]\times 10^{-6}\,{\rm
 GeV}.
 \en
Hence, the weak annihilation ($W$-exchange $E$ or $W$-annihilation
$A$) amplitude has a sizable magnitude comparable to the
color-suppressed internal $W$-emission amplitude $C$ with a large
phase relative to the tree amplitude $T$. As discussed in
\cite{a1a2charm}, it receives long-distance contributions from
nearby resonance via inelastic final-state interactions from the
leading tree or color-suppressed amplitude. The effects of
resonance-induced FSIs can be described in a model independent
manner and are governed by the masses and decay widths of the
nearby resonances. Weak annihilation topologies in $D\to PP$
decays are dominated by nearby scalar resonances via final-state
resacttering. The relative phase between the tree and
color-suppressed amplitudes arises from the final-state
rescattering via quark exchange. This can be evaluated by
considering the $t$-channel chiral-loop contribution or by
applying the Regge pole method (for details, see
\cite{a1a2charm}).

For $D\to SP$ decays, there are several new features. First, one
can have two different external $W$-emission and internal
$W$-emission diagrams, depending on whether the emission particle
is a scalar meson or a pseudoscalar one. We thus denote the prime
amplitudes $T'$ and $C'$ for the case when the scalar meson is an
emitted particle. The quark-diagram amplitudes for various $D\to
SP$ decays are listed in Table IV. Second, because of the
smallness of the decay constant of the scalar meson as discussed
before, it is expected that $|T'|\ll |T|$ and $|C'|\ll |C|$. A
noticeable example is the decay $D^0\to K^-a_0^+$. Its branching
ratio is naively predicted to be of order $10^{-6}$, which is
strongly suppressed compared to the counterpart decay $D^0\to
K^-\pi^+$ in the $PP$ sector. Experimentally, $K^-a_0^+$ has a
branching ratio comparable to $K^-\pi^+$. This implies the
importance of the $W$-exchange term in $D^0\to K^-a_0^+$. Third,
since $K^*_0$ and the light scalars $\sigma,~\kappa,~f_0,~a_0$
fall into two different SU(3) flavor nonets, one cannot apply
SU(3) symmetry to relate the topological amplitudes in $D^+\to
f_0\pi^+$ to, for example, those in $D^+\to \ov K^{*0}_0\pi^+$.
Note that in flavor SU(3) limit, the primed amplitudes $T'$ and
$C'$ diminish in the factorization approach due to the vanishing
decay constants of scalar mesons.

\begin{figure}[t]
\hspace{3cm}
  \psfig{figure=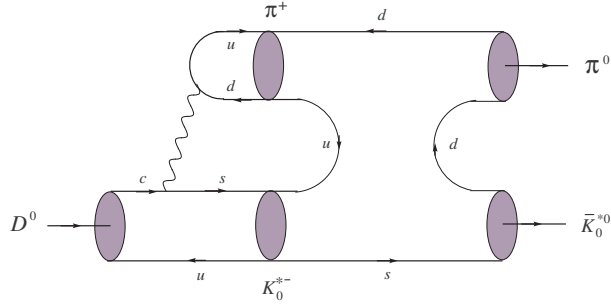,width=10cm}
\vspace{-6cm}
    \caption[]{\small Contributions to $D^0\to \ov K^{*0}_0\pi^0$ from
    the color-allowed weak decay $D^0\to K^{*-}_0\pi^+$ followed by a
    resonant-like rescattering. This has the same topology as the
    $W$-exchange graph.}
\end{figure}

Just as in $D\to PP$ decays, the topological weak annihilation
amplitudes $E$ and $A$, which are naively expected to be helicity
suppressed, can receive large long-distance final-state
interaction contributions. For example, there is a contribution to
the $W$-exchange amplitude $E$ of $D^0\to\ov K^{*0}_0\pi^0$ from
the color-allowed decay $D^0\to K^{*-}_0\pi^+$ followed by a
resonant-like rescattering. As discussed in \cite{a1a2charm}, Fig.
1 manifested at the hadron level receives a $s$-channel resonant
contribution from, for example, the $0^-$ resonance $K(1830)$ and
a $t$-channel contribution with one-particle exchange. Likewise,
the $W$-exchange term in $D^0\to a_0^\pm\pi^\mp$ receives the
$\pi(1800)$ resonance contribution.

\section{Generalized factorization and Analysis}
We will study $D\to SP$ decays within the framework of generalized
factorization in which the hadronic decay amplitude is expressed
in terms of factorizable contributions multiplied by the {\it
universal} (i.e. process independent) effective parameters $a_i$
that are renormalization scale and scheme independent. More
precisely, the weak Hamiltonian has the form
 \be
 H_{\rm eff}={G_F\over\sqrt{2}}V_{cq_1}V_{uq_2}^*\Big[ a_1(\bar uq_2)
 (\bar q_1c)+a_2(\bar q_1 q_2)(\bar uc)\Big]+h.c.,
 \en
with $(\bar q_1q_2)\equiv \bar q_1\gamma_\mu(1-\gamma_5)q_2$. For
hadronic charm decays, we shall use $a_1=1.15$ and $a_2=-0.55$\,.
The parameters $a_1$ and $a_2$ are related to the Wilson
coefficients via
 \begin{eqnarray}  \label{a12}
a_1= c_1(\mu) + c_2(\mu) \left({1\over N_c} +\chi_1(\mu)\right)\,,
\qquad \quad a_2 = c_2(\mu) + c_1(\mu)\left({1\over N_c} +
\chi_2(\mu)\right)\,,
 \end{eqnarray}
where the nonfactorizable terms $\chi_i(\mu)$ will compensate the
scale and scheme dependence of Wilson coefficients $c_i(\mu)$ to
render $a_i$ physical.

\begin{table}[t]
\caption{Quark-diagram amplitudes and  branching ratios for
various $D\to SP$ decays with and without the long-distance weak
annihilation terms induced from final-state interactions. Light
scalar mesons $\sigma,~\kappa,~a_0(980)$ and $f_0(980)$ are
described by the $q^2\bar q^2$ states, while $K^*_0$ is assigned
by $q\bar q$. Experimental results are taken from Table I.
 }
\begin{center}
\begin{tabular}{l l l l l   }
Decay & Amplitude & $\B_{\rm naive}$ & $\B_{\rm FSI}$ & $\B_{\rm expt}$ \\
 \hline
 $D^+\to f_0\pi^+$ & $V_{cd}V_{ud}^*(T+C'+2A)/\sqrt{2}+V_{cs}V_{us}^*\,\sqrt{2}C'$ &
 $3.5\times 10^{-4}$ & see text & $(4.3\pm 1.1)\times 10^{-4}$ \\
 \qquad $\to f_0K^+$ & $V_{cd}V_{us}^*(T+3A)/\sqrt{2}$ & $2.2\times 10^{-5}$ &
 $2.2\times 10^{-5}$ & $(2.0\pm0.5)\times 10^{-4}$ \\
 \qquad $\to a_0^+\ov K^0$ & $V_{cs}V_{ud}^*(T'+C)$ & $1.7\times 10^{-2}$ &
 $1.7\times 10^{-2}$ &  \\
 \qquad $\to a_0^0\pi^+$ & $V_{cd}V_{ud}^*(-T-C')/\sqrt{2}$ &
 $1.7\times 10^{-3}$ & $1.7\times 10^{-3}$ & $(3.2\pm 0.6)\%$ \\
 \qquad $\to \sigma\pi^+$ & $V_{cd}V_{ud}^*(T+C'+2A)$ & input &
  & $(2.1\pm0.5)\times 10^{-3}$ \\
 \qquad $\to\kappa\pi^+$ & $V_{cs}V_{ud}^*(T+C')$ & input & &
 $(6.5\pm1.9)\%$ \\
  \qquad $\to \ov K_0^{*0}\pi^+$ & $V_{cs}V_{ud}^*(T+C')$ & input &
   & $(1.8\pm0.3)\%$ \\
  \hline
  $D^0\to f_0\ov K^0$ & $V_{cs}V_{ud}^*(C+3E)/\sqrt{2}$ & $8.2\times 10^{-4}$
  & input for $E$  & $(6.3\pm1.2)\times 10^{-3}$ \\
  \quad~ $\to a_0^+K^-$ & $V_{cs}V_{ud}^*(T'+E)$ & $2.8\times 10^{-6}$ &
  $1.1\times 10^{-3}$ & $(2.2\pm0.5)\%$ \\
 \quad~ $\to a_0^0\ov K^0$ & $V_{cs}V_{ud}^*(C-E)/\sqrt{2}$ & $3.5\times 10^{-3}$
 & $3.6\times 10^{-3}$ & $(7.9\pm1.7)\%$ \\
  \quad~ $\to a_0^-K^+$ & $V_{cd}V_{us}^*(T+E)$ & $8.1\times 10^{-5}$ & $7.9\times 10^{-5}$ &
  $(2.1\pm1.3)\times 10^{-3}$ \\
  \quad~ $\to a_0^+\pi^-$ & $V_{cd}V_{ud}^*(T'+E)$ & $1.7\times 10^{-7}$ &
  $6.5\times 10^{-5}$ & $(3.4\pm 2.8)\times 10^{-3}$ \\
  \quad~ $\to a_0^-\pi^+$ & $V_{cd}V_{ud}^*(T+E)$ & $1.3\times 10^{-3}$ &
  $1.3\times 10^{-3}$ & $(9.5\pm 7.9)\times 10^{-4}$ \\
  \quad~ $\to K_0^{*-}\pi^+$ & $V_{cs}V_{ud}^*(T+E)$ & $1.3\times 10^{-2}$ &
  $1.1\times 10^{-2}$ &  $(1.18\pm0.25)\%$ \\
   & & & & $(7.0^{+3.1}_{-1.3})\times 10^{-3}$ \\
  \quad~ $\to \ov K_0^{*0}\pi^0$ & $V_{cs}V_{ud}^*(C'+E)/\sqrt{2}$ & $3.9\times 10^{-4}$
  & $3.7\times 10^{-3}$ & $(8.6^{+6.8}_{-2.3})\times 10^{-3}$  \\
  \hline
  $D_s^+\to f_0\pi^+$ & $V_{cs}V_{ud}^*(2T+2A)/\sqrt{2}$ & input &
    & $(1.8\pm0.3)\%$ \\
  \quad~ $\to f_0K^+$ & $V_{cs}V_{us}^*(2T+3A)/\sqrt{2}$ & $1.2\times 10^{-3}$ &
   $1.2\times 10^{-3}$ &  $(1.8\pm0.8)\times 10^{-3}$ \\
  \quad~ $\to \ov K_0^{*0}K^+$ & $V_{cs}V_{ud}^*(C'+A)$ & $4.0\times 10^{-4}$
  & $1.5\times 10^{-3}$ & $(7\pm4)\times 10^{-3}$ \\
  \quad~ $\to K_0^{*0}\pi^+$ & $V_{cd}V_{ud}^*\,T+V_{cs}V_{us}^*\,A$ &
  $1.3\times 10^{-3}$ & $1.1\times 10^{-3}$ & $(2.3\pm1.3)\times 10^{-3}$  \\
\end{tabular}
\end{center}
\end{table}

\subsection{$D\to K^*_0(1430)\pi$}
Among the four measured $D\to K^*_0\pi$ modes: $D^+\to \ov
K^{*0}_0\pi^+$, $D^0\to K^{*-}_0\pi^+,~\ov K^{*0}_0\pi^0$ and
$D_s^+\to K^*_0\pi^+$,  only the first one does not involve weak
annihilation and hence it can be used to fix the form factor for
$D\to K^*_0$ transition. More precisely,
 \be
 A(D^+\to \ov K^{*0}_0\pi^+) &=&
 {G_F\over\sqrt{2}}V_{cs}V_{ud}^*\Big
 [a_1f_\pi(m_D^2-m_{K^*_0}^2)F_0^{DK^*_0}(m_\pi^2)  \non \\
 &+& a_2 f_{K^*_0}(m_D^2-m_\pi^2)F_0^{D\pi}(m_{K^*_0}^2)\Big].
 \en
From a fit to the E791 result $\B(D^+\to \ov
K^{*0}_0\pi^+)=(1.8\pm 0.3)\%$ (see Table I), we obtain
 \be \label{FDK*}
 F_0^{DK^*_0}(0)=1.20\pm 0.07\,.
 \en
As explained in Sec. II, the E791 analysis for $D^+\to
K^-\pi^+\pi^+$ has included the scalar contribution from the
$\kappa$ resonance and found a better improved fit. If the PDG
value of $(3.7\pm 0.4)\%$ for the branching ratio of $D^+\to \ov
K^{*0}_0\pi^+$ is employed \cite{PDG}, we will get a too large
form factor of order 1.60\,.

Under the factorization approximation, the factorizable amplitudes
of other $D\to K^*_0\pi$ decays read
 \be
 A(D^0\to K^{*-}_0\pi^+) &=&
 {G_F\over\sqrt{2}}V_{cs}V_{ud}^*\left[a_1f_\pi(m_D^2-m_{K^*_0}^2)F_0^{DK_0^*}(m_\pi^2)+a_2\la
 K_0^{*-}\pi^+|(\bar s d)|0\ra\la 0|(\bar uc)|D^0\ra \right],  \non \\
 A(D^0\to \ov K_0^{*0}\pi^0) &=& {G_F\over
 2}V_{cs}V_{ud}^* \,a_2\left[f_{K^*_0}(m_D^2-m_\pi^2)F_0^{D\pi}(m^2_{K^*_0})+
 \la  K_0^{*-}\pi^+|(\bar s d)|0\ra\la 0|(\bar uc)|D^0\ra \right],
 \non \\
  A(D^+_s\to K_0^{*0}\pi^+) &=& {G_F\over \sqrt{2}}\,a_1\Big[\,V_{cd}V_{ud}^*
  f_\pi(m_D^2-m_{K^*_0}^2)F_0^{DK^*_0}(m^2_\pi) \non \\ &+&
 V_{cs}V_{us}^*\la K_0^{*0}\pi^+|(\bar u d)|0\ra\la 0|(\bar sc)|D^+_s\ra
 \Big].
 \en
The factorizable (or short-distance) weak annihilation
contribution is conventionally argued to be helicity suppressed.

We see from Table IV that the predictions for $D^0\to
K^{*-}_0\pi^+$ and $D_s^+\to K^{*0}_0\pi^+$ are in agreement with
experiment, whereas $D^0\to \ov K^{*0}_0\pi^0$ and $D_s^+\to \ov
K^{*0}_0K^+$ are too small by an order of magnitude. The latter
implies the importance of long-distance weak annihilation
contributions induced from FSIs. If we assume that the relative
phase and magnitude of the $W$-exchange amplitude $E$ and the
$W$-annihilation amplitude $A$ relative to the external
$W$-emission amplitude $T$ are the same as in the case of $D\to
PP$ decays, namely,
 \be \label{E/T}
 E/T \approx 0.63\,e^{i115^\circ}, \qquad\qquad  A/T\approx
 0.39\,e^{-i65^\circ},
 \en
we find that $D^0\to \ov K^{*0}_0\pi^0$ and $D_s^+\to \ov
K^{*0}_0K^+$ are enhanced by an order of magnitude, while $D^0\to
K^{*-}_0\pi^+$ and $D_s^+\to K^{*0}_0\pi^+$ are affected only
slightly.

\subsection{$D^+\to\kappa\pi^+$}
As noticed in passing, although the decay $D^+\to\kappa\pi^+$ has
very similar topological quark amplitudes as $D^+\to\ov
K^{*0}_0\pi^+$ (see Table IV), they cannot be related to each
other via SU(3) symmetry as $\kappa$ and $K^*_0$ belong to two
different SU(3) flavor nonets. Using the decay constant
$f_{\kappa}=65$ MeV as estimated in Sec. III.B, we find that
$F_0^{D\kappa}(0)=0.85\pm0.10$ from the measured
$D^+\to\kappa\pi^+$ rate. If the decay constant is negligible,
then the form fcator will become smaller,
$F_0^{D\kappa}(0)=0.64\pm0.10$, due to the absence of a
destructive contribution from $C'$.

\subsection{$D^+\to\sigma\pi^+$}
Neglecting $W$-annihilation and taking $m_\sigma=500$ MeV, the
form factor $F_0^{D\sigma}$ extracted from $D^+\to\sigma\pi^+$
reads
 \be
 F_0^{D\sigma}(0)=0.42\pm 0.05\,.
 \en
This is quite different from the fit value $0.8\pm0.2$ obtained in
\cite{Dib} using the E791 data for $D^+\to\pi^+\pi^+\pi^-$
\cite{E791} and the Breit-Wigner description of the $\sigma$
resonance. Note that in the conventional quark model, the $\sigma$
flavor wave function is given by $(u\bar u+d\bar d)/\sqrt{2}$,
while it is $u\bar u d\bar d$ in the four-quark picture.
Therefore, in the SU(3) symmetry limit, the ratio of
$|A(D^+\to\sigma\pi^+)/A(D^+\to\kappa\pi^+)|^2$ is
$|V_{cd}/V_{cs}|^2$ if $\sigma$ is made of four quarks, while it
will be two times smaller if $\sigma$ is a $q\bar q$ state.

\subsection{$D\to f_0(980)\pi,\,f_0(980)K$}

Since the $W$-annihilation contribution is smaller than the
$W$-exchange one in $D\to PP$ decays [see Eq.(\ref{PP})], we will
neglect the $W$-annihilation amplitude $A$ as a first
approximation and determine the $D\to f_0$ form factor from
experiment. We will choose $D_s^+\to f_0\pi^+$ or $D_s^+\to
f_0K^+$ rather than $D^+\to f_0K^+$ to extract $F_0^{Df_0}(0)$.
The reason is as follows. In the SU(3) limit, it is expected that
(see Table IV)
 \be
 {\B(D_s^+\to f_0\pi^+)\over \B(D_s^+\to
 f_0K^+)}=\left|{V_{ud}\over V_{us}}\right|^2.
 \en
It is easily seen that this relation is borne out by experiment.
In contrast, the relation
 \be \label{4quarkrel}
  {\Gamma(D^+\to f_0 K^+)\over \Gamma(D_s^+\to
 f_0K^+)}={1\over 4}\left|{V_{cd}\over V_{cs}}\right|^2
 \en
is different from the experimental ratio which is close to
$|V_{cd}/V_{cs}|^2$. This implies that the decay rate of $D^+\to
f_0K^+$ inferred from FOCUS \cite{FOCUS} is probably too large by
a factor of 4. Indeed, since this mode is Cabibbo doubly
suppressed, it is unlikely that its branching ratio is of the same
order as the Cabibbo singly suppressed one $D^+\to f_0\pi^+$. At
any rate, it is important to check this mode soon. From the decay
$D_s^+\to f_0\pi^+$, we obtain
 \be
 F_0^{D_s^+f_0}(0)=0.52\pm0.04
 \en
and hence $F_0^{Df_0}(0)=0.26\pm0.02$ from Eq. (\ref{Df0}).

Since we have neglected the $W$-annihilation contribution in the
process of extracting the form factor $F_0^{D_s^+f_0}(0)$, we will
consistently ignore this contribution in all $(D,D_s^+)\to
f_0\pi,~f_0K$ decays listed in Table IV. It is clear from this
Table that one needs a sizable $W$-exchange to account for $D^0\to
f_0\ov K^0$. One can utilize this mode to fix the amplitude $E$ to
be
 \be \label{Einf0}
 E/T\approx 0.40\,e^{i100^\circ},
 \en
where we have assumed a phase of $100^\circ$ of the $W$-exchange
term relative to the tree amplitude.

As for the decay $D^+\to f_0\pi^+$, the factorizable internal
$W$-emission amplitude is absent owing to a vanishing $f_0$ decay
constant. Nevertheless, it does receive long-distance contribution
via final-state rescattering, see Fig. 3. Since
$V_{cs}V_{us}^*\approx -V_{cd}V^*_{ud}$ and the amplitude $C'$ is
governed by $a_2$, the amplitude
$V_{cd}V^*_{ud}C'/\sqrt{2}+V_{cs}V^*_{us}\sqrt{2}C'\approx
-V_{cd}V^*_{ud}C'/\sqrt{2}$ will give a constructive contribution
and enhance slightly the decay rate of $D^+\to f_0\pi^+$. At any
rate, the agreement between theory and experiment for this mode
implies that the form factor for $D^+\to f_0$ is indeed smaller
than the one for $D_s^+\to f_0$.

\begin{figure}[t]
\hspace{3cm}
  \psfig{figure=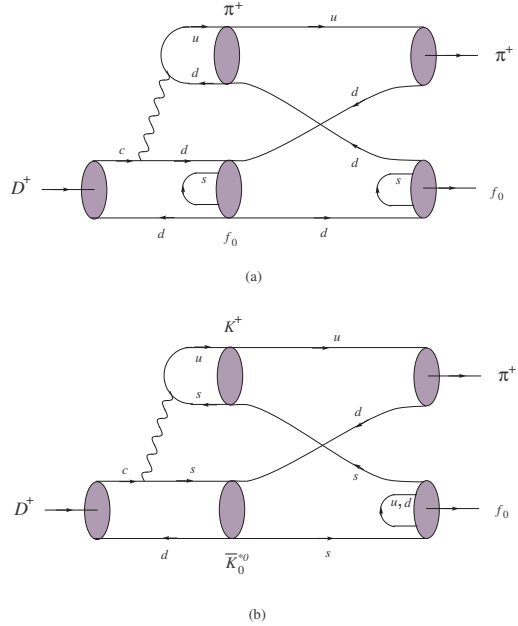,width=8.5cm}
\vspace{0cm}
    \caption[]{\small (a) Long-distance contributions to the amplitude $V_{cd}V_{ud}^*C'$
    of $D^+\to f_0\pi^+$ from the color-allowed weak decay $D^+\to f_0\pi^+$ followed by a
    rescattering via quark exchange, and (b) the long-distance contribution to the amplitude
    $V_{cs}V_{us}^*C'$
    of $D^+\to f_0\pi^+$ from the color-allowed weak decay
    $D^+\to K_0^{*+}\ov K^0,~K^+\ov K^{*0}_0$ followed by a
    rescattering via quark exchange.}
\end{figure}

If $f_0(980)$ is made from $q\bar q$, the ratio of $D^+\to f_0K^+$
to $D_s^+\to f_0K^+$ will be
 \be \label{2quarkrel}
 {\Gamma(D^+\to f_0K^+)\over\Gamma(D_s^+\to
 f_0 K^+)}={1\over 2}\tan^2\theta\left|{V_{cd}\over
 V_{cs}}\right|^2.
 \en
For $\theta\sim 34^\circ$ (see Sec. III.A), it is easily seen that
the two-quark relation (\ref{2quarkrel}) is similar to
(\ref{4quarkrel}) for the four-quark case. However, for
$\theta\sim 140^\circ$ as favored by hadronic $J/\psi$ decays and
the radiative decays $\phi\to f_0(980)\gamma$,
$f_0(980)\to\gamma\gamma$, the $n\bar n$ and $s\bar s$ components
in the $q\bar q$ wave function of $f_0(980)$ have opposite signs.
This means that the interference between the tree amplitude $T$
and the $W$-annihilation amplitude $A$ in the decay, for example,
$D_s^+\to f_0(980)\pi^+$ is opposite in the 2-quark and 4-quark
models. That is, if the interference is constructive in one of the
quark models, it will be destructive in the other model, unless
the relative phase between $T$ and $A$ is $90^\circ$. Therefore,
whether or not one can distinguish between the $q\bar q$ and
$q^2\bar q^2$ pictures for $f_0(980)$ via nonleptonic $D$ decays
depends on the $f_0\!-\!\sigma$ mixing angle and the magnitude and
the phase of the $W$-annihilation term.

Finally we comment on the decays $D^0\to f_0(1370)\ov K^0$,
$D^+\to f_0(1370)\pi^+$ and $D_s^+\to f_0(1370)\pi^+$ which have
been measured by ARGUS \cite{ARGUS}, E687 \cite{E687} and CLEO
\cite{CLEO}, by FOCUS \cite{FOCUS} and by E791 \cite{E791},
respectively, with the results
 \be
 \B(D^0\to f_0(1370)\ov K^0)\B(f_0(1370)\to\pi^+\pi^-) &=& \cases{
 (4.7\pm1.4)\times 10^{-3} & ARGUS,E687 \cr (5.9^{+1.8}_{-2.7})\times 10^{-3}, & CLEO} \non \\
 \B(D^+\to f_0(1370)\pi^+)\B(f_0(1370)\to K^+K^-) &=&
 (6.2\pm 1.1)\times 10^{-4}\qquad{\rm FOCUS}  \non \\
 \B(D^+_s\to f_0(1370)\pi^+)\B(f_0(1370)\to\pi^+\pi^-) &=&
 (3.3\pm 1.2)\times 10^{-3}\qquad{\rm E791}
 \en
Since the branching fractions of $f_0(1370)\to\pi^+\pi^-,K^+K^-$
are unknown, the individual branching ratio of $D$ decays into
$f_0(1370)$ cannot be determined at present. Nevertheless, if
$f_0(1370)$ is a $n\bar n$ state in nature, the decay $D^+_s\to
f_0(1370)\pi^+$ can only proceed through the topological
$W$-annihilation diagram. Hence, this will be the first direct
evidence for a non-vanishing $W$-annihilation amplitude $A$ in
$D\to SP$ decays. The other modes $D^0\to f_0(1370)\ov K^0$ and
$D^+\to f_0(1370)\pi^+$ proceed via the internal $W$-emission
diagram $C$ and the external $W$-emission diagram $T$,
respectively. Taking into account the phase-space correction and
noting that $D^+\to f_0(1370)\pi^+$ is Cabibbo singly suppressed
while the other two are Cabibbo allowed, it is obvious that
$|T|>|C|> |A|$, as it should be.

\subsection{$D\to a_0(980)\pi,\,a_0(980)K$}
Because the primed amplitudes $T'$ and $C'$ are largely suppressed
relative to the unprimed ones $T$ and $C$ owing to the smallness
of the $a_0$ decay constant, it is interesting to notice that the
neutral state $a_0^0\ov K^0$ is not color suppressed relative to
the charged mode $a_0^+K^-$, contrary to the case of
$D^0\to\pi^+K^-,\pi^0\ov K^0$. It is also anticipated that
$a_0^-\pi^+\gg a_0^+\pi^-$, a relation that cannot be tested  by
the present preliminary data as they do not have enough
statistical significance.

Just as the $D$ decays to $\sigma$, $\kappa$, $f_0(980)$ and
$K^*_0(1430)$, one may use the channel $D^+\to a_0^0\pi^+$ to fix
the form factor for $D\to a_0$ transition. However, in the SU(3)
limit, one has the relations (see Table IV)
 \be
  R_1\equiv {\B(D^+\to a_0^0\pi^+)\over \B(D^+\to
 \kappa\pi^+)}={1\over 2}\left|{V_{cd}\over V_{cs}}\right|^2, \qquad
 R_2\equiv {\B(D^+\to a_0^0\pi^+)\over \B(D^+\to
 \sigma\pi^+)}\approx {1\over 2}.
 \en
Experimentally, $R_1\sim {1\over 2}$ and $R_2\sim 15$. This
indicates that the preliminary data of $D^+\to a_0^0\pi^+$ is too
large by at least an order of magnitude. Therefore, we will
instead use the value of 0.77 for $F_0^{Da_0}(0)$ (see Sec.
III.C). For the $W$-exchange amplitude we can apply Eq.
(\ref{Einf0}).

The results of calculations are shown in Table IV. Obviously all
the predicted branching ratios for $D\to a_0\pi,~a_0K$ (except for
$D^0\to a_0^-\pi^+$) are too small by one to two orders of
magnitude when compared with experiment. Note that $D^0\to
a_0^-K^+$ is Cabibbo doubly suppressed and it appears to be very
unlikely that it has a large branching ratio of order $10^{-3}$.
From Table IV we also see that $D^+\to a_0^+\ov K^0$ has the
largest branching ratio among the two-body decays $D\to
a_0\pi(K)$.

If we fit the $D\to a_0$ form factor to $D^+\to a_0^0\pi^+$, we
will get $F_0^{Da_0}(0)=3.4$ and the large discrepancy between
theory and experiment will be greatly improved. However, in the
meantime we also predict that $\B(D^+\to a_0^+\ov K^0)=0.35$ which
is obviously too large. Moreover, it is impossible to achieve this
abnormally large $D\to a_0$ form factor in the quark model.

It is possible that one has to apply the Breit-Wigner
approximation for $a_0(980)$ to derive the branching ratios for
$D\to a_0\pi,~a_0K$ from the three-body decays of charmed mesons.
Furthermore, the fraction of $a_0(980)\to K\ov K$ should be pinned
down. It will be interesting to compare the experimental results
with the predictions exhibited in Table IV.

\section{Conclusions}

The nonleptonic weak decays of charmed mesons into a scalar meson
and a pseudoscalar meson are studied. The scalar mesons under
consideration are $\sigma$ [or $f_0(600)$], $\kappa$, $f_0(980)$,
$a_0(980)$ and $K^*_0(1430)$. The main conclusions are:

 \begin{enumerate}
 \item Studies of the mass
spectrum of scalar mesons and their strong as well as
electromagnetic decays suggest that the light scalars below or
near 1 GeV form an SU(3) flavor nonet and are predominately the
$q^2\bar q^2$ states, while the scalar mesons above 1 GeV can be
described as a $q\bar q$ nonet with a possible mixing with $0^+$
$q\bar q$ and glueball states. Therefore, we designate $q^2\bar
q^2$ to $\sigma,~\kappa,~a_0(980),~f_0(980)$ and $q\bar q$ to
$K^*_0$.

 \item The topological quark-diagram scheme for $D\to SP$ decays is more
complicated than the case of $D\to PP$. One can have two different
external $W$-emission and internal $W$-emission diagrams,
depending on whether the emission particle is a scalar meson or a
pseudoscalar one. The quark-diagram amplitude for the case when
the emitted particle is a scalar meson is largely suppressed
relative to the one when the pseudoscalar meson is emitted.
Moreover, the former amplitude vanishes in the limit of SU(3)
symmetry.

 \item The charmed meson to $\kappa$ and $K_0^*$ transition form
factors are extracted from the Cabibbo-allowed decays
$D^+\to\kappa\pi^+,~\ov K^{*0}_0\pi^+$, respectively, while
$D^+\to\sigma$ and $D_s^+\to f_0$ ones are inferred from
$D^+\to\sigma\pi^+$ and $D_s^+\to f_0\pi^+$, respectively, based
on the assumption of negligible $W$ annihilation. We show that a
large form factor for $D\to K^*_0$ is expected. The relation
$F_0^{D^+f_0}=F_0^{D_s^+f_0}/2$ obtained in the 4-quark picture
for $f_0(980)$ leads to a prediction for $D^+\to f_0\pi^+$ in
agreement with experiment. Note that the value of the form factor
$F_0^{D\sigma}(0)=0.42\pm0.05$ obtained in this work is very
different from the one $0.8\pm0.2$ quoted in the literature. It is
pointed out that the ISGW model and its improved version the ISGW2
model predict too small $D\to S$ form factors even at zero recoil.

 \item Except for the Cabibbo doubly suppressed decay $D^+\to f_0K^+$, the
data of $D\to\sigma\pi,~ f_0\pi,~f_0K,~K^*_0\pi$ can be
accommodated  in the generalized factorization approach. Sizable
weak annihilation contributions induced from final-state
interactions are crucial for understanding the data. For example,
the importance of the $W$-exchange term is implied by the decays
$D^0\to f_0\ov K^0,~\ov K^{*0}_0\pi^0$ and the $W$-annihilation
one by $D_s^+\to\ov K^{*0}_0K^+$. Without $W$-exchange or
$W$-annihilation contributions, the decay rates of these modes
will be too small by one order of magnitude. The branching ratio
of $D^+\to f_0K^+$ is predicted to be of order $10^{-5}$ and
should be tested soon.

 \item The predicted $D\to a_0\pi,~a_0K$ rates are too small by one to two orders of
magnitude when compared with the preliminary measurements. It is
pointed out that $D^+\to a_0^+\ov K^0$ should have the largest
branching ratio among the decays $D\to a_0\pi,~a_0K$.

\item If $f_0(980)$ is a $q\bar q$ state in nature, it must
contain not only $s\bar s$ but also $u\bar u$ and $d\bar d$
content. The $f_0\!-\!\sigma$ mixing angles inferred from the
hadronic decays $J/\psi\to f_0\phi/\omega$, the radiative decay
$\phi\to f_0\gamma$ followed by $f_0\to\gamma\gamma$, and the
strong coupling of $f_0$ to $K\ov K$ and $\pi\pi$ are not quite
compatible with each other. If $\theta\sim 140^\circ$, then it
will be possible to distinguish between the two-quark and
four-quark pictures for $f_0(980)$ in the decay, for example,
$D_s^+\to f_0\pi^+$. In the SU(3) symmetry limit, the ratio of
$|A(D^+\to\sigma\pi^+)/A(D^+\to\kappa\pi^+)|^2$ can be different
by a factor of 2 in these two different pictures.

 \end{enumerate}

\vskip 2.5cm \acknowledgments This work was supported in part by
the National Science Council of R.O.C. under Grant No.
NSC91-2112-M-001-038.


\end{document}